\documentstyle[epsf,12pt,mycite,mydina4p]{article}
\newcommand{\mayor}{\mbox{\raisebox{-0.4ex}
{$\;\stackrel{>}{\scriptstyle \sim}\;$}}}
\newcommand{\menor}{\mbox{\raisebox{-0.4ex}
{$\;\stackrel{<}{\scriptstyle \sim}\;$}}}

\def\etjet{E_T^{jet}}
\def\etajet{\eta^{jet}}
\def\phijet{\varphi^{jet}}
\def\etcal{E_{T,cal}^{jet}}
\def\etacal{\eta_{cal}^{jet}}
\def\phical{\varphi_{cal}^{jet}}

\def\wcal{W^{cal}}

\def\etaphi{\eta-\varphi}

\def\rs{R_{SEP}}

\def\etar{-1<\etajet<2}

\def\wr{134 $<W<$ 277 GeV}

\def\wrg{134 $<W<$ 190 GeV}
\def\wrgg{190 $<W<$ 233 GeV}
\def\wrggg{233 $<W<$ 277 GeV}

\def\seta{d\sigma/d\etajet}

\def\q2{Q^2}
\def\pb1{pb$^{-1}$}
\def\gp{\gamma p}
\def\g2{GeV$^2$}

\def\rr1{R=1.0}
\def\r7{R=0.7}
\def\R71{R=1.0\ {\rm and}\ 0.7}
\def\RR5{R=0.5}
\def\RRR{R=1.0,\ 0.7\ {\rm and}\ 0.5}

\def\sigr{\sigma(R)}

\def\epr{e^+p\rightarrow e^+\ +\ {\rm jet}\ + \ {\rm X}}

\def\rtr{< \hspace{-0.2cm} r_{tracks} \hspace{-0.2cm} >}
\def\rtd{< \hspace{-0.17cm} r_{dijet}  \hspace{-0.17cm} >}

\begin{document}

\vspace{1 cm}

\begin{titlepage}

\title{
{\bf High-$E_T$ Inclusive Jet Cross Sections\\
        in Photoproduction at HERA}
\author{ZEUS Collaboration}}

\date{}
\maketitle

\vspace{5 cm}

\begin{abstract}

Inclusive jet differential cross sections for the reaction $\epr$ with
quasi-real photons have been measured with the ZEUS detector at HERA. These
cross sections are given for the photon-proton centre-of-mass energy interval
\wr\ and jet pseudorapidity in the range $\etar$ in the laboratory frame. The
results are presented for three cone radii in the $\etaphi$ plane, $\RRR$.
Measurements of $\seta$ above various jet-transverse-energy thresholds up to
25~GeV and in three ranges of $W$ are presented and compared to
next-to-leading order (NLO) QCD calculations. For jets defined with $\rr1$
differences between data and NLO calculations are seen at high $\etajet$ and
low $\etjet$. The measured cross sections for jets defined with $\r7$ are well
described by the calculations in the entire measured range of $\etajet$ and
$\etjet$. The inclusive jet cross section for $\etjet>21$~GeV is consistent
with an approximately linear variation with the cone radius $R$ in the range
between 0.5 and 1.0, and with NLO calculations.

\end{abstract}

\vspace{-18cm}
{\noindent
 DESY 98-018 \newline
 February 1998}

\setcounter{page}{0}
\thispagestyle{empty}
\pagenumbering{Roman}
\def\3{\ss}
\parindent0.cm
\parskip 3mm plus 2mm minus 2mm

\newpage

\begin{center}
{\Large  The ZEUS Collaboration}
\end{center}

  J.~Breitweg,
  M.~Derrick,
  D.~Krakauer,
  S.~Magill,
  D.~Mikunas,
  B.~Musgrave,
  J.~Repond,
  R.~Stanek,
  R.L.~Talaga,
  R.~Yoshida,
  H.~Zhang  \\
 {\it Argonne National Laboratory, Argonne, IL, USA}~$^{p}$
% ...

\newpage

% ...
$^{  19}$ now at ATLAS Collaboration, Univ. of Munich \\
% ...
$^{  30}$ partially supported by the Foundation for German-Russian
Collaboration DFG-RFBR \\
\hspace*{3.5mm} (grant no. 436 RUS 113/248/3 and no. 436 RUS 113/248/2)\\
% ...

\newpage

\begin{tabular}[h]{rp{14cm}}

$^{a}$ &  supported by the Natural Sciences and Engineering Research
          Council of Canada (NSERC)  \\
$^{b}$ &  supported by the FCAR of Qu\'ebec, Canada  \\
$^{c}$ &  supported by the German Federal Ministry for Education and
          Science, Research and Technology (BMBF), under contract
          numbers 057BN19P, 057FR19P, 057HH19P, 057HH29P \\
$^{d}$ &  supported by the MINERVA Gesellschaft f\"ur Forschung GmbH,
          the German Israeli Foundation, and the U.S.-Israel Binational
          Science Foundation \\
$^{e}$ &  supported by the German-Israeli Foundation, the Israel Science
          Foundation, the U.S.-Israel Binational Science Foundation, and by
          the Israel Ministry of Science \\
$^{f}$ &  supported by the Italian National Institute for Nuclear Physics
          (INFN) \\
$^{g}$ &  supported by the Japanese Ministry of Education, Science and
          Culture (the Monbusho) and its grants for Scientific Research \\
$^{h}$ &  supported by the Korean Ministry of Education and Korea Science
          and Engineering Foundation  \\
$^{i}$ &  supported by the Netherlands Foundation for Research on
          Matter (FOM) \\
$^{j}$ &  supported by the Polish State Committee for Scientific
          Research, grant No.~115/E-343/SPUB/P03/002/97, 2P03B10512,
          2P03B10612, 2P03B14212, 2P03B10412 \\
$^{k}$ &  supported by the Polish State Committee for Scientific
          Research (grant No. 2P03B08308) and Foundation for
          Polish-German Collaboration  \\
$^{l}$ &  partially supported by the German Federal Ministry for
          Education and Science, Research and Technology (BMBF)  \\
$^{m}$ &  supported by the Fund for Fundamental Research of Russian Ministry
          for Science and Edu\-cation and by the German Federal Ministry for
          Education and Science, Research and Technology (BMBF) \\
$^{n}$ &  supported by the Spanish Ministry of Education
          and Science through funds provided by CICYT \\
$^{o}$ &  supported by the Particle Physics and
          Astronomy Research Council \\
$^{p}$ &  supported by the US Department of Energy \\
$^{q}$ &  supported by the US National Science Foundation \\

\end{tabular}

\end{titlepage}

\newpage
\parindent 5mm
\parskip 0mm
\pagenumbering{arabic}
\setcounter{page}{1}
\normalsize

\section{Introduction}

 At HERA, photon-proton reactions are studied by means of $ep$ scattering at
low four-momentum transfers squared ($\q2\approx 0$). In photoproduction, two
types of QCD processes contribute to the production of jets \cite{owens,drees}
at leading order (LO): either the photon interacts directly with a parton in
the proton (the direct process) or the photon acts as a source of partons
which interact with those in the proton (the resolved process). Differential
cross sections for inclusive jet photoproduction using a cone algorithm have
been previously presented as a function of the jet pseudorapidity\footnote{The
ZEUS coordinate system is defined as right-handed with the $Z$ axis pointing
in the proton beam direction, hereafter referred to as forward, and the $X$
axis horizontal, pointing towards the centre of HERA. The pseudorapidity is
defined as $\eta=-\ln(\tan\frac{\theta}{2})$, where the polar angle $\theta$
is taken with respect to the proton beam direction.} ($\etajet$) and
transverse energy ($\etjet$) for $\etjet$ up to 17~GeV
\cite{h193,zeoct94,h196}. The calculated cross sections depend on the proton
parton distributions in the region of Bjorken-$x$ above approximately
$10^{-2}$, where they are well constrained by other measurements \cite{sfss}.
Such jet measurements therefore offer a potential means of studying the parton
distributions in the photon
\cite{drees,muchos,kramer2,kramer3,harris1,stefano1} at higher scales than those
probed in $e^+e^-$ interactions \cite{oldee}. However, various aspects of the
comparison between theory and experiment need to be addressed before a reliable
determination of the photon parton distributions can be made.

 Next-to-leading order (NLO) calculations including resolved plus direct
processes and using the NLO parametrisations of the photon parton
distributions of GRV \cite{grv} have been compared \cite{kramer2}
to our previous measurements \cite{zeoct94}. Discrepancies were observed in
the forward region ($\etajet>1$) for low $\etjet$ ($\etjet\sim 8$~GeV)
which prevented any strong conclusion being drawn on the photon parton
distributions. Moreover, the comparison of the transverse energy flow between
data and leading-logarithm parton-shower Monte Carlo simulations \cite{zeoct94}
showed a discrepancy at high $\etajet$ that could be attributed to energy not
associated with the hard-scattering process (the `underlying event'). Such an
underlying event is not included in the NLO calculations and, therefore, the
comparison between data and NLO calculations becomes problematic. The
transverse energy inside the cone of the jet in the $\etaphi$ plane due to the
underlying event is naively expected to be proportional to the area covered by
the cone. Therefore, measurements performed with different cone radii can
elucidate the effects of a possible underlying event. In addition, the NLO
calculations for jets defined with a cone radius $R\approx 0.7$ are expected
to be most stable with respect to variations of the renormalisation and
factorisation scales \cite{sdellis}. Measurements of the jet cross sections in
different ranges of the $\gp$ centre-of-mass energy ($W$) provide a further
means of comparing data and calculations.

 In this paper, measurements of $\seta$ are presented for various cone radii.
In each case, measurements of $\seta$ integrated above four different $\etjet$
thresholds (14, 17, 21 and 25 GeV) are performed. For $\rr1$, the measurement
of $\seta$ is extended to higher $\etjet$ values as compared to the previous
data \cite{zeoct94}. We have, in addition, a better understanding of the energy
scale of the jets. First measurements of $\seta$ in three regions of $W$ are
presented for $\R71$. The dependence of the inclusive jet cross section on the
jet cone radius is presented. NLO calculations \cite{kramer3,harris1} which
include resolved plus direct processes are compared to the measurements.

 The data sample used in this analysis was collected with the ZEUS detector in
$e^+p$ interactions at the HERA collider and corresponds to an integrated
luminosity of 2.65~\pb1, which is a five-fold increase in statistics over the
previous analysis \cite{zeoct94}.

\section{Experimental conditions}

 During 1994 HERA operated with protons of energy $E_p=820$~GeV and positrons
of energy $E_e=27.5$~GeV. The ZEUS detector is described in detail in
\cite{sigtot,status}. The main subdetectors used in the present analysis are
the central tracking system positioned in a 1.43~T solenoidal magnetic field
and the uranium-scintillator sampling calorimeter (CAL). The tracking system
was used to establish an interaction vertex and to cross-check the energy
scale of the CAL. Energy deposits in the CAL were used in the jet finding and
to measure jet energies. The CAL is hermetic and consists of 5918~cells each
read out by two photomultiplier tubes. Under test beam conditions, the CAL has
energy resolutions of 18\%/$\sqrt{E}$ for electrons and 35\%/$\sqrt{E}$ for
hadrons. Jet energies are corrected for the energy lost in inactive material
in front of the CAL which is typically about one radiation length (see
Section~5). The effects of uranium noise were minimised by discarding cells in
the inner (electromagnetic) or outer (hadronic) sections if they had energy
deposits of less than 60~MeV or 110~MeV, respectively. The luminosity was
measured from the rate of the brems\-strahlung process
$e^+p\rightarrow e^+p\gamma$. A three-level trigger was used to select events
online \cite{status,jetshape}.

\section{Data selection and jet search}

 Offline, events from quasi-real photon-proton collisions were selected using
similar criteria as employed previously \cite{zeoct94}. The main steps are
briefly discussed here. The contamination from beam-gas interactions, cosmic
showers and beam-halo muons is negligible after demanding: a) at least two
tracks pointing to the vertex; b) the vertex position along the beam axis to
lie in the range $-29<Z<36$~cm; c) fewer than five tracks not
associated with the vertex and compatible with an interaction upstream in the
direction of the proton beam; and d) the number of tracks not associated to
the vertex be less than 10\% of the total number of tracks. Deep-inelastic
(DIS) charged-current $e^+p$ scattering events are rejected by requiring the
total missing transverse momentum (${p_T\hspace{-3.5mm}\slash\hspace{1.5mm}}$)
to be small compared to the total transverse energy ($E^{tot}_T$):
${p_T\hspace{-3.5mm}\slash\hspace{1.5mm}} /\sqrt{E^{tot}_T} 
<2$~GeV$^{\frac{1}{2}}$. DIS neutral-current events with an identified
scattered positron candidate in the CAL, according to the algorithm described
in \cite{zenov93}, are removed from the sample. The selected sample consists of
events from $e^+p$ interactions with $\q2\menor 4$ \g2\ and a median of
$\q2\approx 10^{-3}$ \g2. The events are restricted to the kinematic range \wr\
using the procedure described in Section~5.

 An iterative cone algorithm in the $\etaphi$ plane \cite{cone2,snow} is used 
to reconstruct jets from the energy measured in the CAL cells. A detailed
description of the algorithm can be found in \cite{jetshape}. The jets
reconstructed from the CAL cell energies are called $cal$ jets and the
variables associated with them are denoted by $\etcal$, $\etacal$ and
$\phical$. The axis of the jet is defined according to the Snowmass convention
\cite{snow}, where $\etacal$ ($\phical$) is the transverse-energy weighted
mean pseudorapidity (azimuth) of all the CAL cells belonging to that jet.
Events with at least one jet satisfying $\etcal>10$~GeV and $-1<\etacal<2$ are
retained. Three samples of jets have been selected depending on the cone
radius used in the jet search: 18897 jets for $\rr1$, 11197 jets for $\r7$
and 7070 jets for $\RR5$. The only significant remaining background is from
unidentified DIS neutral current interactions with $\q2>$~4 \g2, which is
estimated using Monte Carlo techniques to be below 2\%.

\section{Monte Carlo simulation}

 Samples of events were generated using Monte Carlo (MC) simulations to
determine the response of the detector to jets of hadrons and the correction
factors for the inclusive jet cross sections.

 The programs PYTHIA~5.7~\cite{pythia} and HERWIG~5.8~\cite{herwig} were used
to generate photoproduction events for resolved and direct processes. In PYTHIA
the positron-photon vertex was modelled according to the
Weizs\"{a}cker-Williams approximation. In the case of HERWIG, the exact matrix
elements were used for direct processes ($e^+g\rightarrow e^+q\bar q$ and
$e^+q\rightarrow e^+qg$) and the equivalent photon approximation for resolved
processes. Events were generated using GRV-HO \cite{grv} for the photon parton
distributions and MRSA \cite{mrsa} for the proton parton distributions. In
addition, samples of events using the LAC1 parametrisation \cite{lac1} for the
photon parton distributions were considered. In both generators, the partonic
processes were simulated using LO matrix elements, with the inclusion of
initial and final state parton showers. Fragmentation into hadrons was
performed using the LUND \cite{lund} string model as implemented in JETSET
\cite{jetset} in the case of PYTHIA, and the cluster model \cite{webber} in the
case of HERWIG. Samples of events were generated with different values of
the cut-off on the transverse momentum of the two outgoing partons starting at
$\hat p_{Tmin}= 8$~GeV. For the measurements presented in this paper, the
events generated using the PYTHIA and HERWIG programs have been used for
calculating energy corrections and for correcting the data for detector and
acceptance effects. The corrections provided by the PYTHIA generator have been
used as default values and the ones given by the HERWIG generator have been
used to estimate the systematic errors coming from the fragmentation model.
 
 Additional samples of events were generated using the option of multiparton
interactions (MI) in PYTHIA. This option, which applies only to resolved 
processes, adds interactions between the partons in the proton and the 
photon remnants to the hard scattering process of the event. These multiparton
interactions are calculated as LO QCD processes and give an estimation of the
underlying event. The PYTHIA MI events were generated with a cut-off for
the effective minimum transverse momentum for multiparton interactions of
1~GeV \cite{jetshape} and with a cut-off on the transverse momentum of the two
outgoing partons from the hard scattering of $\hat p_{Tmin}= 8$~GeV.

 All generated events were passed through the ZEUS detector and trigger
simulation programs \cite{status}. They were reconstructed and analysed 
by the same program chain as the data.

 For the Monte Carlo events, the jet search is performed from the energy
measured in the CAL cells in the same way as in the data. The same jet
algorithm is also applied to the final state particles. In this search, all
particles with lifetimes longer than $10^{-13}$~s and with polar angles
between $5^{\circ}$ and $175^{\circ}$ are considered. The jets found are
called $hadron$ jets and the variables associated with them are denoted by
$E^{jet}_{T,had}$, $\eta^{jet}_{had}$, and $\varphi^{jet}_{had}$. $Hadron$
jets with $E^{jet}_{T,had}>14$ GeV and $-1<\eta^{jet}_{had}<2$ are selected.

\section{Energy corrections}

 The fivefold increase in statistics in 1994 allowed the CAL energy scale to
be studied in more detail than in ref. \cite{zeoct94}. The comparison of the
energy measured in the central region of the CAL to the momentum measured in
the tracking system for the scattered positron in neutral current DIS events,
and the transverse momentum balance in neutral current DIS events, showed a
($6\pm 3$)\% difference between data and MC \cite{zef294}. This 6\% disagreement
has been corrected for in the present analysis. In the analysis of the 1993
data, the possibility of such a discrepancy was allowed for in the
systematic uncertainties.

 Particles impinging on the CAL lose energy in the inactive material in front
of the CAL. The inactive material constitutes about one radiation length
except in the region around the rear beampipe, $\theta\mayor 170^{\circ}$, and
the support structures, $25^{\circ}\menor\theta\menor 45^{\circ}$ and
$130^{\circ}\menor\theta\menor 145^{\circ}$, where it reaches $2.5$ radiation
lengths. For the measurements presented here, the transverse energy of the
jets has also been corrected for these energy losses as explained below.

 The comparison of the reconstructed jet variables between the $hadron$ and 
the $cal$ jets in simulated events shows no significant systematic shift in
the angular variables $\etacal$ and $\phical$ with respect to $\etajet_{had}$
and $\phijet_{had}$. Therefore, no correction is needed for $\etajet$ and
$\phijet$ ($\etajet\approx\etacal$ and $\phijet \approx \phical$). However,
the transverse energy of the $cal$ jet underestimates that of the $hadron$ jet
by an average amount of 16\% with an r.m.s. of 11\%. The transverse energy
corrections to $cal$ jets averaged over the azimuthal angle were determined
using the MC events. These corrections are constructed as multiplicative
factors, $C(\etcal,\etacal)$, which, when applied to the $E_T$ of the $cal$
jets provide the `corrected' transverse energies of the jets,
$\etjet=C(\etcal,\etacal) \times \etcal$.

 The method of Jacquet-Blondel \cite{jacblo}, applied to the photoproduction
regime \cite{jorges}, is used to estimate $W$ from the energies measured in the
CAL cells: $\wcal=\sqrt{2 E_p \cdot (E-p_Z)}$, where $E$ is the total CAL
energy and $p_Z$ is the $Z$ component of the directed energy measured in the
CAL cells. Due to energy lost in the inactive material in front of the CAL and
to particles lost in the rear beampipe, $\wcal$ systematically underestimates
$W$ by approximately 10\% with an r.m.s. of 5\%. This effect is adequately
reproduced by the MC simulation of the detector. To compensate for this
underestimation, MC samples of events were used to determine a correction
procedure to $\wcal$ as a function of $\wcal$ and of the pseudorapidity of the
most backward jet in the event ($\etajet_{min}$). This correction has been
constructed as a multiplicative function, $Y(\wcal,\etajet_{min})$, in a
similar way as the correction to the jet transverse energy. When applying the
function $Y$ to $\wcal$, $W=Y(\wcal,\etajet_{min})\times\wcal$, the corrected
$\gamma p$ centre-of-mass energy is obtained, and events with \wr\ are retained.

\begin{figure}
\epsfysize=18cm
\centerline{\epsffile{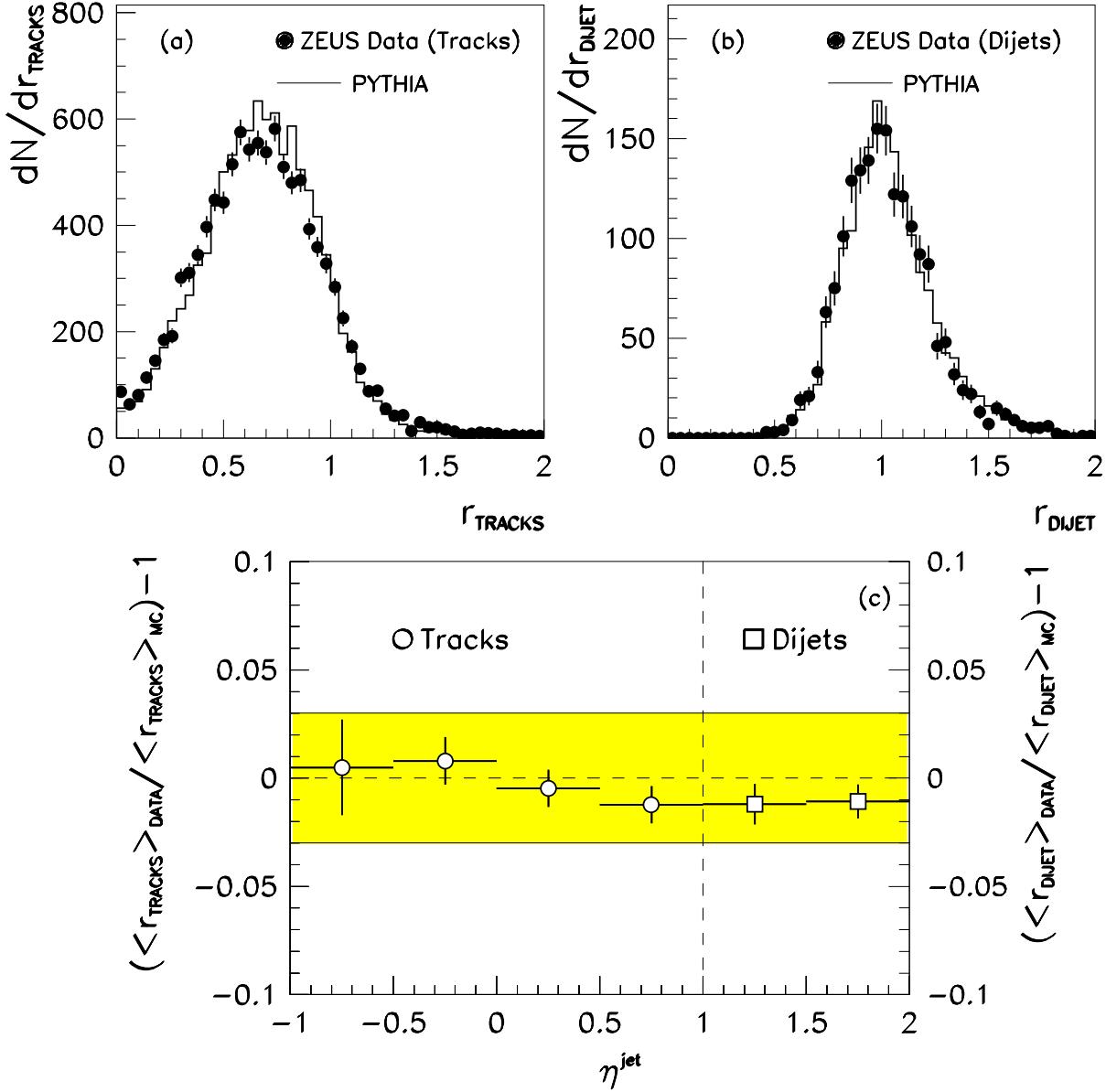}}
\caption{\label{inclusive370}
{(a) The distribution of $r_{tracks}\equiv 
E^{jet}_{T,tracks}/E^{jet}_{T,cal}$ for the inclusive jet data sample 
with $\rr1$ (black dots) and as reproduced by the PYTHIA generator and 
detector simulation (histogram, normalised to the number of jets in the 
data); (b) the distribution of $r_{dijet}\equiv 
E^{jet}_{T,cal}$(forward)$/E^{jet}_{T,cal}$(central) for the
dijet data sample (one jet in the forward region and the other in the
central region) with $\rr1$ (black dots) and as reproduced by the
PYTHIA generator and detector simulation (histogram, normalised to the number
of jets in the data); (c) the values of the quantity
$(\rtr_{data}/\rtr_{MC})-1$ (circles) and
$(\rtd_{data}/\rtd_{MC})-1$ (squares). The shaded region displays
the band of $\pm 3$\% around zero.}}
\end{figure}

 The response of the CAL to jets has been checked by the following procedure
\cite{thesis}. In the central region ($|\etajet|<1$), the multiplicity
distribution and the $p_T$-spectrum of charged particles within the $cal$ jets
have been compared for data and Monte Carlo samples using the reconstructed
tracks. The tracks were required to be in  the ranges $|\eta^{track}|<1.5$ and
$p^{track}_T>300$~MeV, where $p^{track}_T$ is the transverse momentum of the
track with respect to the beam axis. Tracks were associated with a $cal$ jet
when the extrapolated trajectory reached the CAL within the cone of the $cal$
jet. PYTHIA describes well all the measured distributions. In this $\etajet$
region, the momenta of the tracks in the $cal$ jet are used to determine the
total transverse energy carried by the charged particles,
$E^{jet}_{T,tracks}$. Then, the ratio
$r_{tracks}\equiv E^{jet}_{T,tracks}/\etcal$ is formed, and the distributions
of this ratio for the inclusive $cal$ jet sample with $\rr1$ in data and
simulations are compared, as shown in Figure~\ref{inclusive370}a. The mean value
of the distribution in $r_{tracks}$ has been determined as a function of
$\etajet$ for data ($\rtr_{data}$) and simulations ($\rtr_{MC}$). From the
values of the quantity $(\rtr_{data}/\rtr_{MC})-1$, shown in
Figure~\ref{inclusive370}c (circles), we conclude that the energy scale of
the jets with $|\etajet|<1$ is correct to within $\pm 3$\%.

 In the forward region, $1<\etajet<2$, the energy scale of the jets is studied
using the transverse energy imbalance in dijet events with one jet in the
central region and the other in the forward region. The distributions of the
ratio $r_{dijet}\equiv\etcal$(forward jet)$/\etcal$(central jet) in data and
simulations are compared in Figure~\ref{inclusive370}b. The values of the
quantity $(\rtd_{data}/\rtd_{MC})-1$ (see Figure~\ref{inclusive370}c, square
symbols) show that in the forward region the energy scale of the jets is also
correct to within $\pm 3$\%. 

 It is noted that since the widths of the $r_{tracks}$ and $r_{dijet}$
distributions in the data are reasonably well described by the PYTHIA
simulations, the resolution in the energy of the jets is also correctly
described.

 This procedure has been also applied to the inclusive $cal$ jet sample with
$\r7$ and leads to the same conclusions. The use of HERWIG instead of PYTHIA
gives similar results. Therefore, a $\pm 3$\% uncertainty on the energy
scale of the jets is included as a systematic variation in the present
analysis.

\section{Jet profiles}

 The presence of energy not associated to the hard-scattering process (the
`underlying event') in the data has been investigated through the study of the
transverse energy flow around the jet axis both inside and outside of the jet
cone.

\begin{figure}
\epsfysize=18cm
\centerline{\epsffile{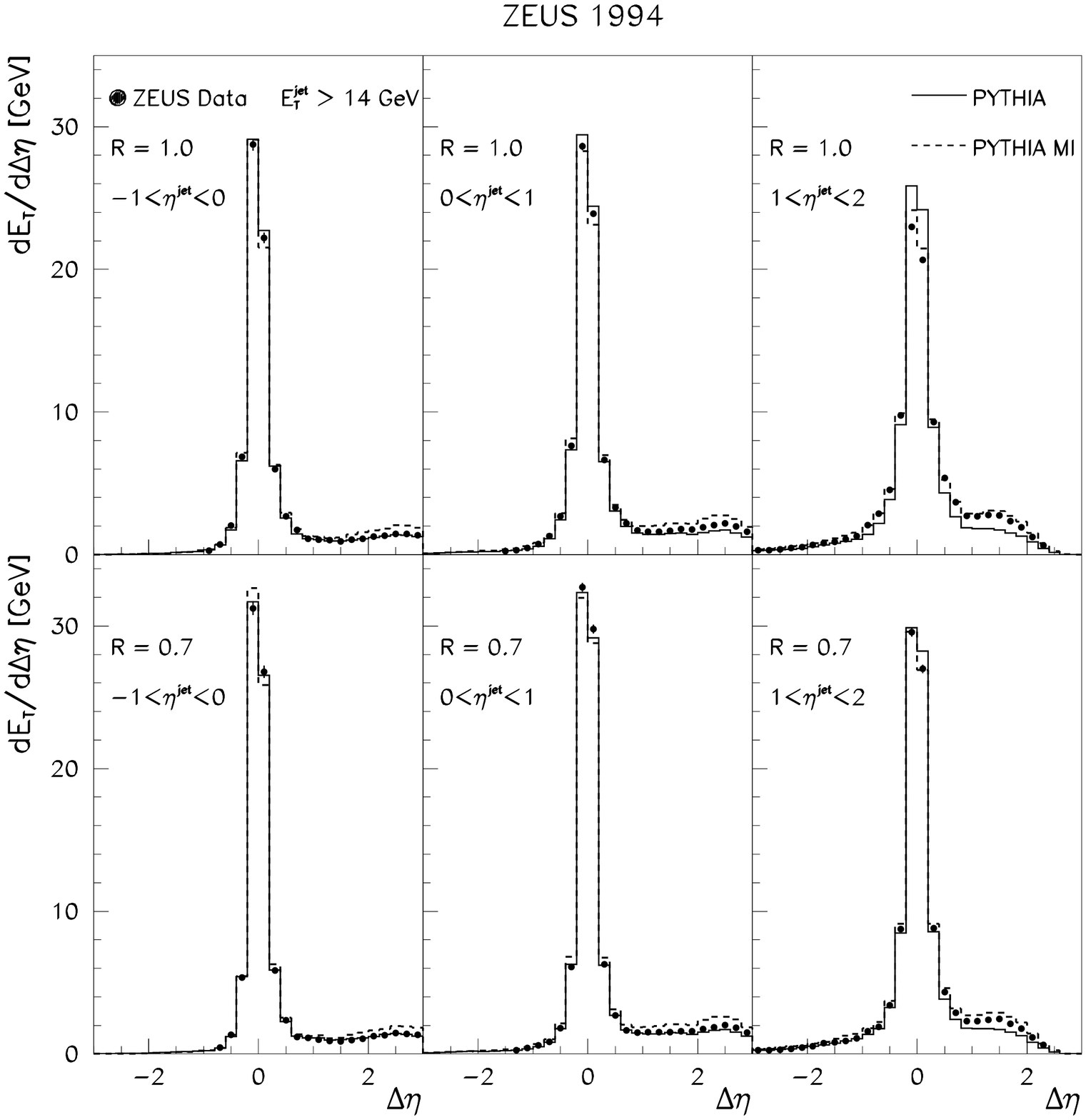}}
\caption{\label{inclusive271}
{Uncorrected transverse energy profiles for jets with $\etjet>14$~GeV and
$\R71$ as a function of the distance from the jet axis, $\Delta\eta$
(integrated over $|\Delta\varphi| < \pi/2$), in three $\etajet$ regions
(black dots). For comparison, PYTHIA and PYTHIA MI simulations including
resolved plus direct processes are shown as the solid and dashed histograms,
respectively.}}
\end{figure}

\begin{figure}
\epsfysize=18cm
\centerline{\epsffile{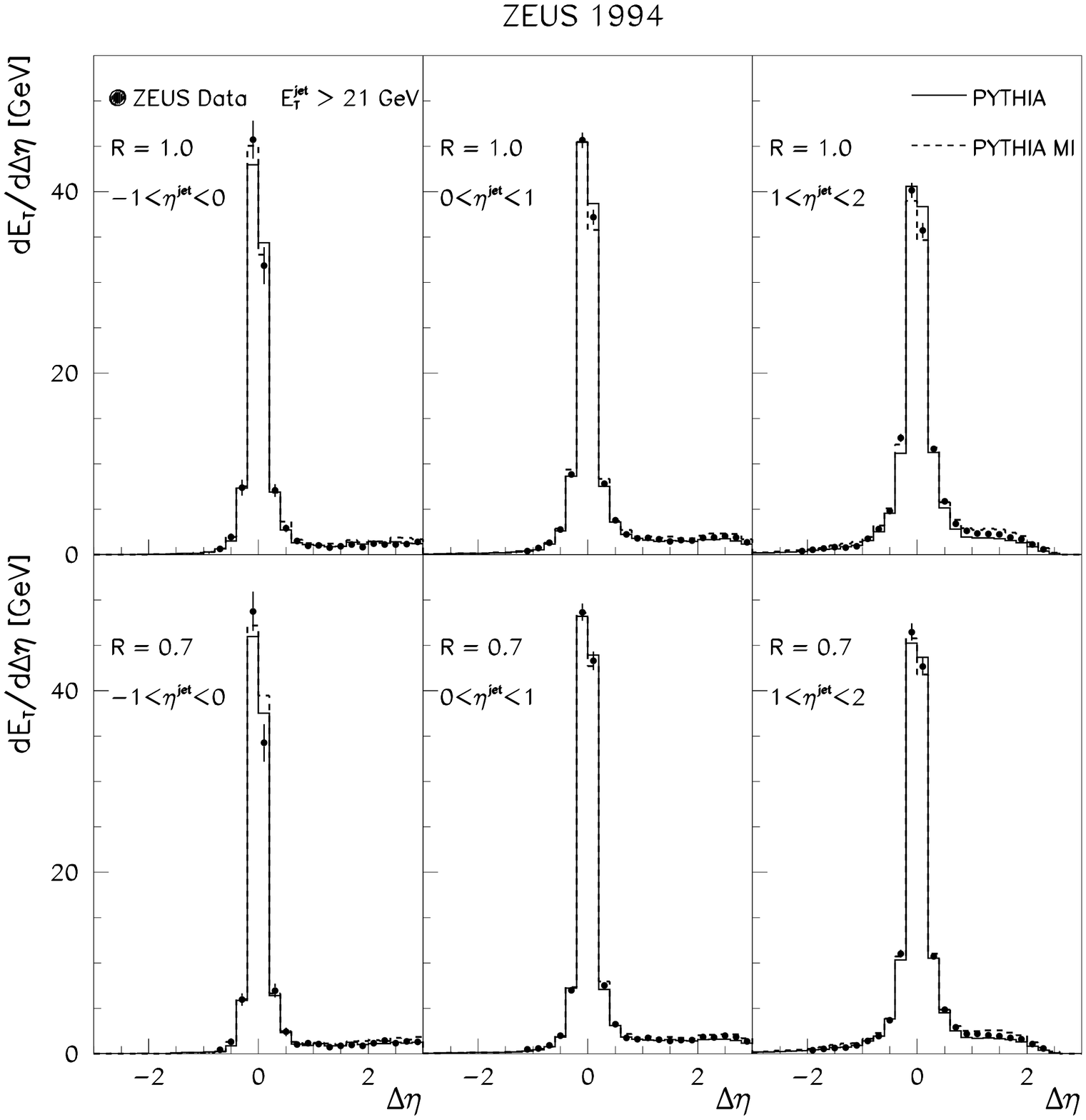}}
\caption{\label{inclusive272}
{Uncorrected transverse energy profiles for jets with $\etjet>21$~GeV and
$\R71$ as a function of the distance from the jet axis, $\Delta\eta$
(integrated over $|\Delta\varphi| < \pi/2$), in three $\etajet$ regions
(black dots). For comparison, PYTHIA and PYTHIA MI simulations including
resolved plus direct processes are shown as the solid and dashed histograms,
respectively.}}
\end{figure}

 The transverse energy profile around the jet axis was measured using the
energies and angles of the CAL cells uncorrected for detector effects. The
distribution of transverse energy in the hemisphere of the jet, as a function
of $\Delta\eta \equiv \eta_{cell} - \etajet$ and integrated over
$|\Delta\varphi| \equiv |\varphi_{cell} - \phijet|<\pi/2$,
is shown in Figures~\ref{inclusive271} and \ref{inclusive272} for the inclusive
jet data samples ($\R71$) in three $\etajet$ ranges and two $\etjet$
regions\footnote{The decrease of the $\Delta\eta$ distribution seen both in
data and the simulations in the region $\Delta\eta > 2$ for the forward
jets ($1<\etajet<2$) is a geometric effect: the most forward edge of the CAL 
is at $\eta = 4.3$.}. The data exhibit a pronounced peak at $\Delta\eta=0$ and
an asymmetric pedestal. The height of the peak increases as $\etjet$
increases. As a function of $\etajet$, it is fairly constant in the region
$-1<\etajet<1$ and decreases in the region $\etajet>1$; this decrease is most
significant for jets with $\rr1$. The height of the pedestal for
$\Delta\eta>1$ (proton side) slightly increases with increasing $\etajet$, the
effect being more pronounced for $\rr1$ and low $\etjet$.

 The expectations from PYTHIA simulations including resolved plus direct
processes are compared to the data in Figures~\ref{inclusive271} and
\ref{inclusive272}. The transverse energy profile in the data is well described
by the simulations of PYTHIA except for jets with $\etajet>1$ and lowest
$\etjet$ ($\etjet\approx 14$~GeV). In this region, an excess of transverse
energy outside of the jet cone with respect to PYTHIA simulations is observed
\cite{h193,zeoct94,h196,zedij95}. The excess is reduced for jets defined with
$\r7$ in comparison to jets defined with $\rr1$ (see Figure~\ref{inclusive271}).
In order to simulate an increased energy flow, the PYTHIA MI generator is used,
which gives rise to energy not associated with the hard-scattering process.
PYTHIA MI gives an improved description of the
data for forward low-$\etjet$ jets with $\rr1$, but lies above the data for
$\etajet<1$ in the case of $\rr1$ and in all $\etajet$ ranges for $\r7$. For
jets with $\etjet>21$~GeV, no significant discrepancies are observed between
data and PYTHIA simulations with or without multiparton interactions (see
Figure~\ref{inclusive272}).

 The internal structure of the jets may be investigated using the jet shape,
defined as the average fraction of the jet transverse energy that occurs inside
an inner cone concentric with the jet defining cone \cite{sdellis}. The shape
of jets selected using $\rr1$ has been recently measured in photoproduction at
HERA \cite{jetshape} and found to be well described by the PYTHIA calculations
except for the inclusive production of jets with $\etajet>1$ and low $\etjet$
(14~GeV~$<\etjet<17$~GeV). We have performed the same type of analysis for
jets with $\r7$ and, in this case, the measured jet shapes (not shown) are
well described by the PYTHIA (with or without multiparton interactions)
calculations in the entire $\etajet$ region.

 These observations indicate that the uncertainties on the jet measurements
due to possible underlying event contributions become reduced at high $\etjet$
($\etjet>21$~GeV) or when using a reduced cone radius ($\r7$).

\section{Acceptance corrections and systematic uncertainties}

 The MC generated event samples of resolved and direct processes were used to
compute the acceptance corrections to the inclusive jet distributions. These
correction factors take into account the efficiency of the trigger, the
selection criteria and the purity and efficiency of the jet reconstruction.
The differential cross sections $\seta$ are then obtained by applying
bin-by-bin corrections to the measured jet distributions. The predictions of
the generators PYTHIA and HERWIG for the uncorrected distributions were
compared to the data for several choices of the parton densities in the photon
and proton and for various combinations of resolved and direct processes. A
good description of the $\etajet$ data distributions is obtained by the MC
except for forward low-$\etjet$ jets with $\rr1$. The bin-by-bin correction
factors lie between 0.7 and 1.4 depending on $\etajet$, $\etjet$ threshold and
$W$ region considered. The dominant effect arises from migrations over the
$\etjet$ threshold.

 A detailed study of the sources contributing to the systematic uncertainties
of the measurements has been performed. The study of the systematic 
uncertainties includes (a typical value for each item is indicated):
\begin{itemize}
 \item Use of the HERWIG generator to evaluate the energy corrections to
        $cal$ jets and the correction factors to the observed inclusive jet
        distributions. The effect of this variation is typically within
        $\pm 5$\% in the region $0.5<\etajet<2$ and increases to $\approx 10$\%
        for $\etajet<0.5$.
 \item Uncertainties in the simulation of the trigger and the variation of
       the cuts used to select the data within the ranges allowed by the
       comparison between data and MC simulations ($\approx 5$\%).
 \item Use of the PYTHIA generator including multiparton interactions in
        resolved processes to evaluate the energy corrections to
        $cal$ jets and the correction factors to the observed inclusive jet
        distributions ($\approx 3$\%). In the region of forward low-$\etjet$
        jets with $\rr1$, an improved description of the data is obtained by
        using PYTHIA MI.
 \item Choice of different parton densities in the photon (GRV-HO and LAC1)
        for the generation of the PYTHIA MC samples ($\approx 2$\%).
\end{itemize}
All these systematic uncertainties have been added in quadrature to the
statistical errors and are shown as thin error bars in the figures.

\begin{itemize}
 \item The absolute energy scale of the $cal$ jets in simulated events has been
varied by $\pm 3$\% for the reasons discussed in Section~5. The effect of this
variation on the inclusive jet cross sections is $\approx\pm 12$\% in the
region $0<\etajet<2$, and increases up to $\approx 35$\% for
$\etajet\approx -1$. This uncertainty represents the dominant source of
systematic error and is highly correlated between measurements at different
$\etajet$ points. It is shown as a shaded band in each figure.
\end{itemize}

 In addition, there is an overall normalisation uncertainty of 1.5\% from the
luminosity determination, which is not included.

\section{Results}

\subsection{Differential cross sections}
 
\begin{figure}
\epsfysize=18cm
\centerline{\epsffile{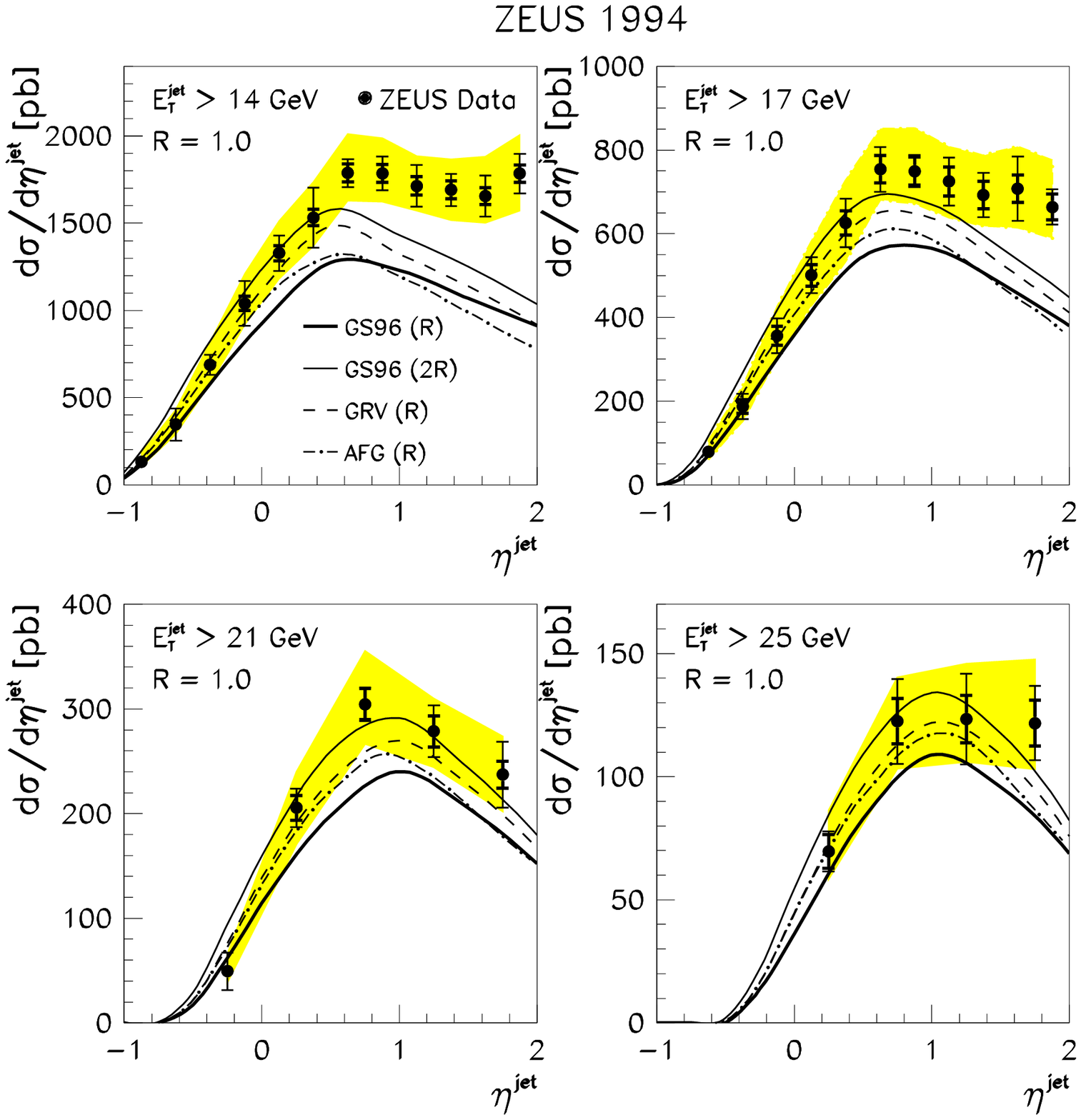}}
\caption{\label{inclusive191}{Differential $e^+p$ cross section
$\seta$ for inclusive jet production integrated above $\etjet$ from four
different thresholds ($\etjet>$ 14, 17, 21 and 25~GeV) in the kinematic region
defined by $\q2\leq 4$~\g2\ and \wr\ for jets with a cone radius $\rr1$. The
thick error bars represent the statistical errors of the data, and the thin
error bars show the statistical errors and systematic uncertainties $-$not
associated with the absolute energy scale of the jets$-$ added in quadrature.
The shaded bands display the uncertainty due to the absolute energy scale of
the jets. For comparison, NLO calculations for three parametrisations of the
photon parton distributions, $\mu=\etjet$ and for two different values of
$\rs$ are shown: AFG $\rs=R$ (dot-dashed line), GRV-HO $\rs=R$ (dashed line),
GS96 $\rs=R$ (thick solid line) and GS96 $\rs=2R$ (thin solid line). The
values of $\rs$ used are indicated in parentheses. In all cases, the CTEQ4M
proton parton distributions have been used.}}
\end{figure}

\begin{figure}
\epsfysize=18cm
\centerline{\epsffile{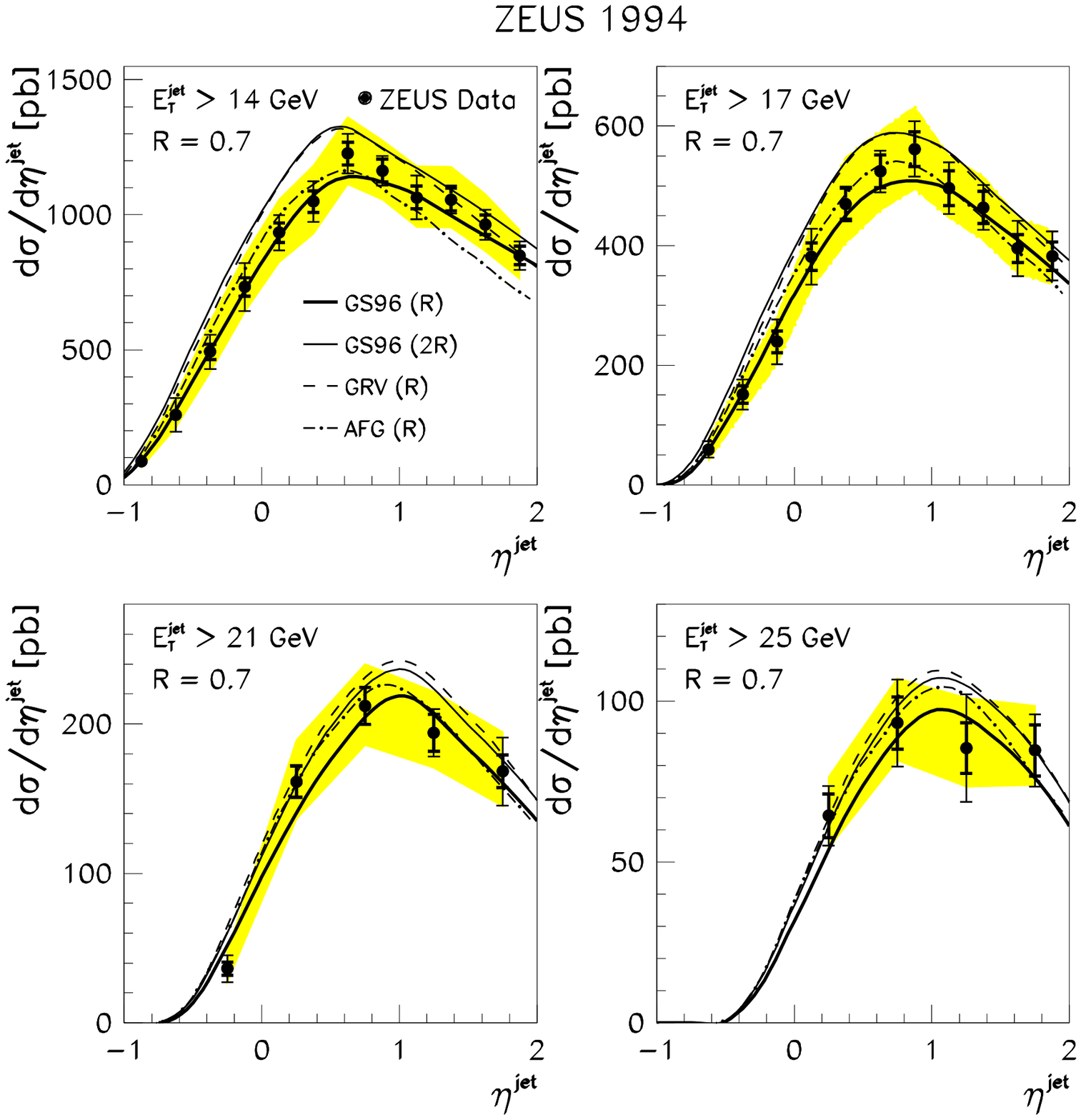}}
\caption{\label{inclusive192}{Differential $e^+p$ cross section $\seta$ for
inclusive jet production for jets with a cone radius $\r7$. Other details as
in Figure~\ref{inclusive191}.}}
\end{figure}

\begin{figure}
\epsfysize=18cm
\centerline{\epsffile{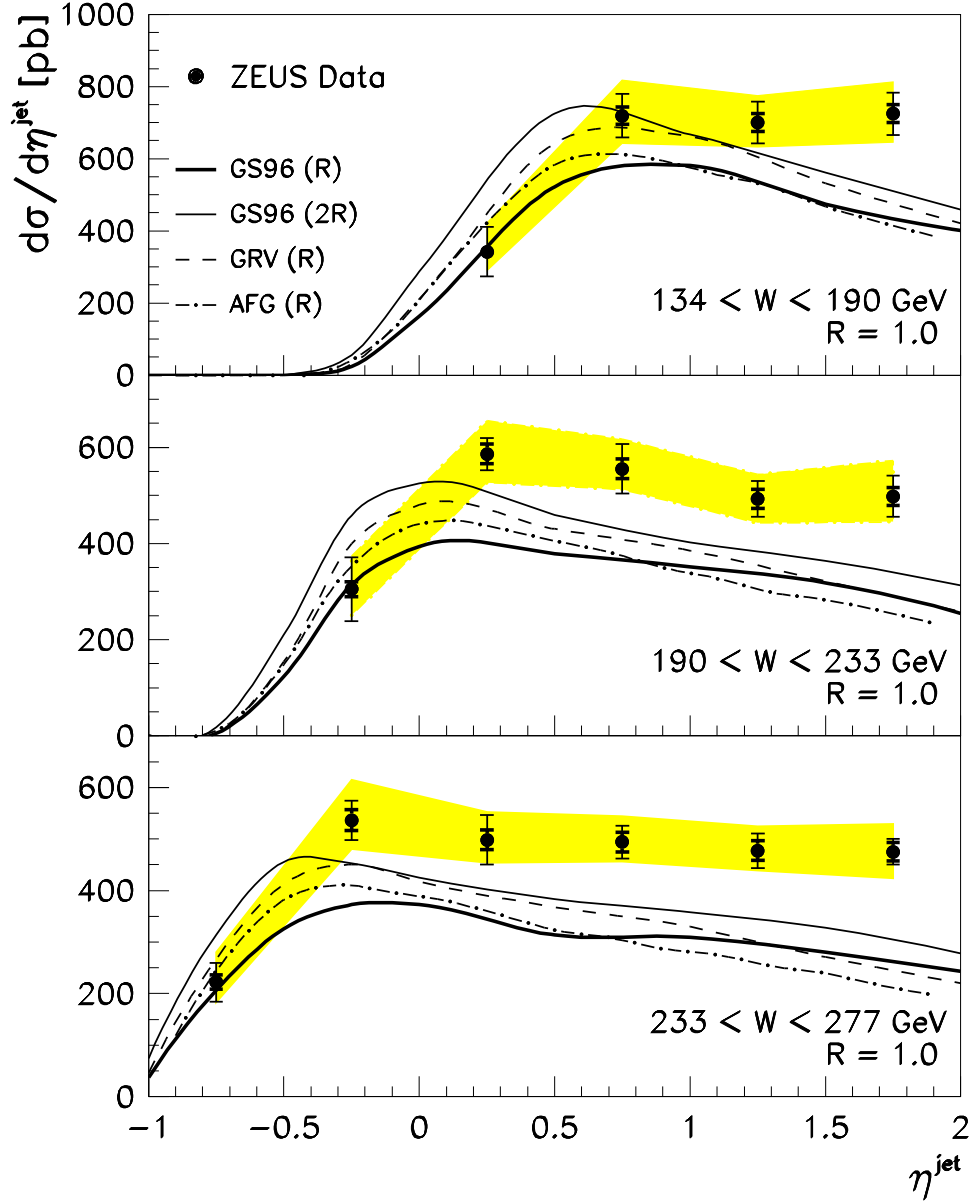}}
\vspace{-0.5cm}
\caption{\label{inclusive201}{Differential $e^+p$ cross section $\seta$ for
inclusive jet production integrated above $\etjet>14$~GeV in the kinematic
region defined by $\q2\leq 4$~\g2\ and in three regions of $W$: \wrg\ (upper
plot), \wrgg\ (middle plot) and \wrggg\ (lower plot) for jets with a cone radius
$\rr1$. Other details as in Figure~\ref{inclusive191}.}}
\end{figure}

\begin{figure}
\epsfysize=18cm
\centerline{\epsffile{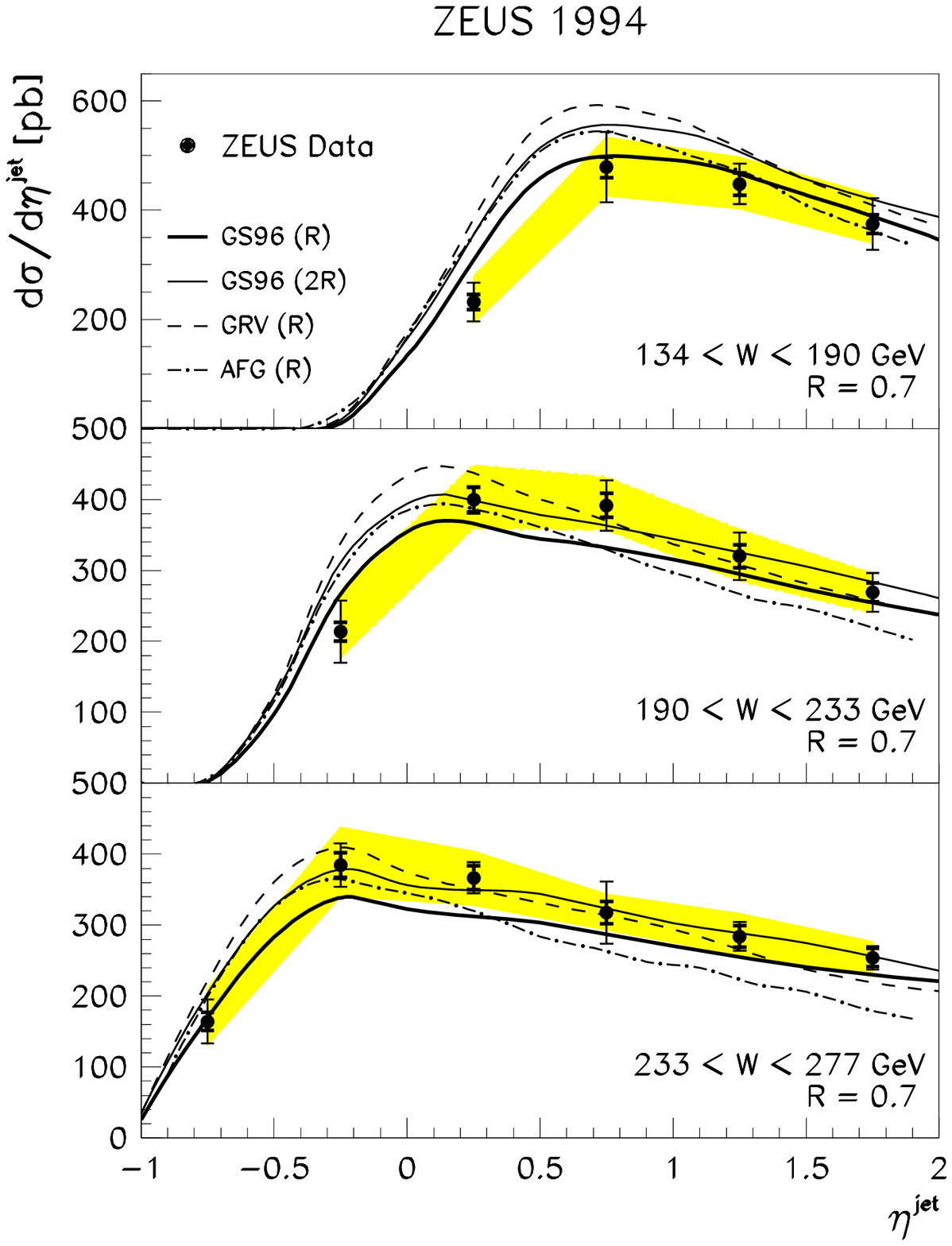}}
\caption{\label{inclusive202}{Differential $e^+p$ cross section $\seta$ for
inclusive jet production for jets with a cone radius $\r7$. Other details as
in Figure~\ref{inclusive201}.}}
\end{figure}

 We present measurements of inclusive differential jet cross sections for the
reaction
$$\epr$$
in the kinematic region defined by $\q2\leq 4$~\g2\ and \wr . These cross
sections refer to jets at the hadron level with cone radii of $\R71$ units in
the $\etaphi$ plane. The cross section $\seta$ has been measured in the
$\etajet$ range between $-1$ and $2$ integrated above $\etjet$ from four
different thresholds ($\etjet>$ 14, 17, 21 and 25~GeV). The cross section
$\seta$ for $\etjet>14$~GeV has also been measured for three different regions
of $W$: \wrg , \wrgg\ and \wrggg . The results are presented in
Figures~\ref{inclusive191} to \ref{inclusive202} and in Tables~\ref{tabsec1} to
\ref{tabsec4}.

 For $\etjet>14$ and 17~GeV, the behaviour of the cross section as a function
of $\etajet$ in the region $\etajet>1$ is very different for $\rr1$ and $\r7$
(see Figures~\ref{inclusive191} and \ref{inclusive192}): it is constant for
$\rr1$ whereas it decreases as $\etajet$ increases for $\r7$. On the other
hand, the behaviour for $\etjet>21$ and 25~GeV is approximately the same in
both $\R71$ cases. There are two effects which contribute to the observed
differences in the region $\etajet>1$ for $\etjet>14$ and 17~GeV: a) the jets
become broader as $\etajet$ ($\etjet$) increases (decreases) \cite{jetshape}
and, b) the pedestal in the jet profile is integrated over approximately half
the area for jets with $\r7$. In addition, the height of the pedestal 
(see Section~6) is larger for forward jets with $\rr1$. Therefore, the
differences between the cross sections for the two radii can be attributed to
the fact that the use of $\r7$ selects more collimated jets and suppresses
the underlying event contribution.

 The results for $\seta$ in different regions of $W$ for $\etjet >14$~GeV and
with $\rr1$ $(\r7)$ are presented in Figure~\ref{inclusive201} 
(\ref{inclusive202}). For $\rr1$, the cross section increases with increasing 
values of $\etajet$ and is constant in the high $\etajet$ region, whereas 
for $\r7$ the cross section decreases as $\etajet$ increases in the high 
$\etajet$ region. For increasing values of $W$ the maximum of the cross 
section with $\rr1$ $(\r7)$ shifts to lower values of $\etajet$. As the energy
of the incoming quasi-real photon increases, $W$ increases and the events are
boosted more backwards in the laboratory frame.

\subsection{Comparison to NLO calculations}
 
 NLO QCD calculations of $\seta$ \cite{kramer3,harris1} are compared to our 
measurements in Figures~\ref{inclusive191} to~\ref{inclusive202}. These 
predictions include resolved and direct processes. The CTEQ4M \cite{cteq4} 
proton parton densities have been used. For the photon parton distributions, 
the AFG \cite{afg}, GRV-HO \cite{grv} and GS96 \cite{gs} parametrisations have
been used\footnote{The calculations using GRV-HO or GS96 are from \cite{kramer3}
and those using AFG from \cite{harris1}. For the same photon parton
distributions, the calculations from \cite{kramer3} and \cite{harris1} differ
typically by less than $\pm 5$\%.}. In the calculations shown here, the
renormalisation and factorisation scales have been chosen equal to $\etjet$
and $\alpha_s$ was calculated at two loops with
$\Lambda^{(4)}_{\overline{MS}}=296$~MeV \cite{kramer3}.

 The comparison of the data with NLO calculations is subject to the
uncertainty in matching the experimental and theoretical jet algorithms. Since
the calculations include only up to three partons in the final state, the
maximum number of partons in a single jet is two. Therefore, the overlapping
and merging effects of the experimental jet algorithm are not reproduced in
the theoretical calculation \cite{sdellis,giele}. An attempt was made to
simulate these effects by introducing an ad-hoc $\rs$ parameter \cite{sdellis}:
two partons are not merged into a single jet if their separation in the
$\etaphi$ plane is more than $\rs$. The calculations of the cross sections
shown in Figures~\ref{inclusive191} to \ref{inclusive202} have been made for
$\rs=R$. In addition, the calculations using GS96 and $\rs=2R$ are also shown.
The spread of the calculations using GS96 for $\rs=R$ and $\rs=2R$ indicates
the magnitude of the theoretical uncertainty due to these effects.

 As discussed above, the NLO calculations refer to jets built out of at most
two partons whereas the measurements refer to jets at the hadron level. An
estimate of the effects of hadronisation has been obtained by comparing the
cross sections for jets of hadrons and jets of partons calculated with the
PYTHIA generator. The ratio of ($\seta$[hadrons])/($\seta$[partons]) for jets
with $\rr1$ ($\r7$) is relatively constant as a function of $\etajet$ and
within approximately 10\% (20\%) of unity. Due to the approximations used in
the MC simulations, these estimations are not to be taken as corrections to
the parton level for the measurements presented here.

 The NLO calculations give a good description of the measured differential
cross sections in magnitude and shape for $\etjet>$ 21 and 25~GeV for both 
cone radii $\R71$. For $\etjet>$ 14~GeV, the behaviour of the measured cross
sections is different for $\R71$, whereas the calculations exhibit the same
shape for both radii. For $\rr1$, the shape of the cross section is well
described for $-1<\etajet<0.5$. For higher values of $\etajet$, the measured
cross section is constant, as discussed in Section~8.1, whereas the 
theoretical curves decrease. These differences are not present when $\r7$ is 
used: the NLO calculations describe well the magnitude and shape of the 
measured differential cross sections with $\r7$ for all $\etjet$ thresholds 
in the entire range of $\etajet$.

 The measured differential cross sections in ranges of $W$ for $\rr1$ and
$\etjet>$ 14 GeV are reasonably well described for low values of $\etajet$,
whereas the NLO calculations fail to describe the high $\etajet$ region. The
excess of the measured cross section with respect to the calculations
increases with increasing $W$. On the other hand, the measured differential
cross sections in bins of $W$ are reasonably well described by the NLO 
calculations using $\r7$ in the entire region of $\etajet$.

 The failure of the NLO calculations to describe the measured cross section 
for forward low-$\etjet$ jets with $\rr1$ may be due to the following
effects: a) the uncertainty due to the choice of renormalisation and
factorisation scales is larger than in the case $\r7$ (see next section), and
b) non-perturbative contributions like that of the underlying event, which is
reduced for jets with $\r7$, are not included. On the other hand, for jets
defined with $\r7$ the measured cross sections are well described by the
calculations and the uncertainties on the measurements are comparable to the
spread of the predictions using different parametrisations of the photon
parton distributions. 

\subsection{Cone radius dependence of the cross section}

\begin{figure}
\epsfysize=18cm
\centerline{\epsffile{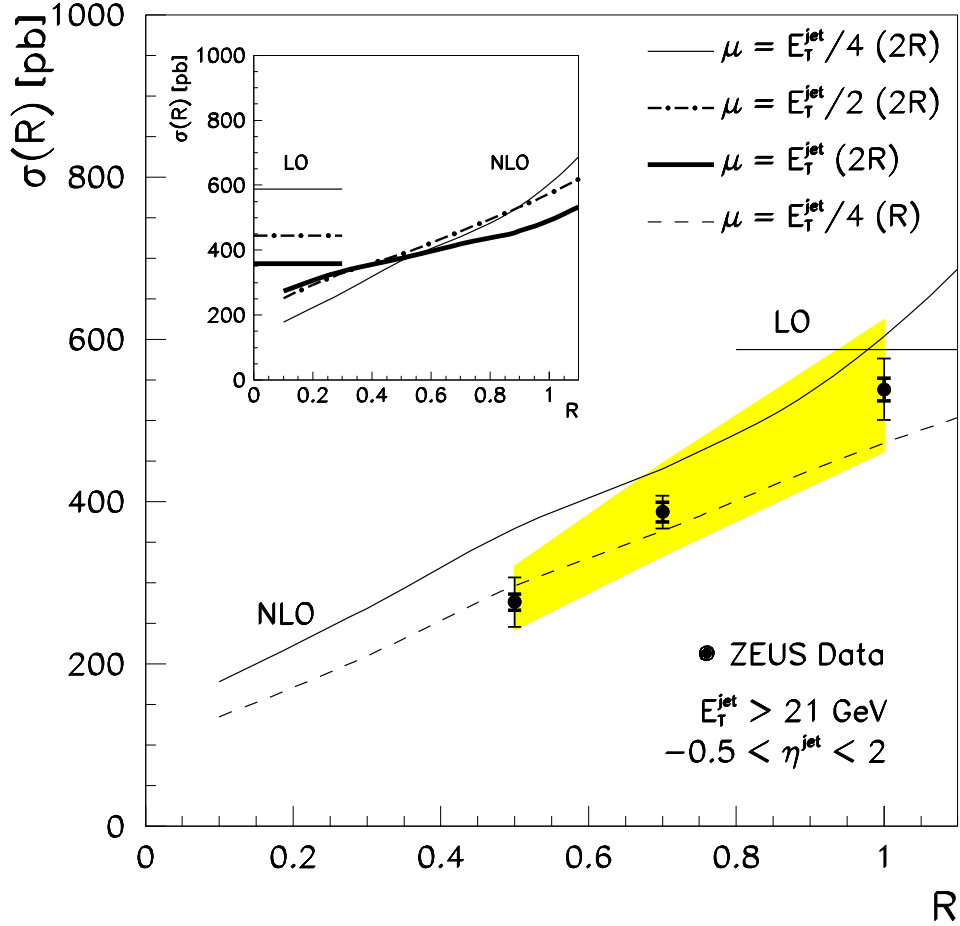}}
\caption{\label{inclusive350}{$e^+p$ cross section $\sigr$ as a function of the
jet cone radius $R$ for inclusive jet production integrated above 
$\etjet>21$~GeV and $-0.5<\etajet <2$ in the kinematic region defined by 
$\q2\leq 4$~\g2\ and \wr . The thick error bars represent the statistical 
errors of the data, and the thin error bars show the statistical errors and 
systematic uncertainties $-$not associated with the absolute energy scale of 
the jets$-$ added in quadrature. The shaded band displays the uncertainty 
due to the absolute energy scale of the jets. LO and NLO calculations using
the GS96 (CTEQ4) parametrisations of the photon (proton) parton distributions
and $\mu=\etjet/4$ for two choices of the parameter $\rs$ are shown. The
values of $\rs$ used are indicated in parentheses. The inset shows the
calculations for a fixed value of $\rs=2R$ and various choices of $\mu$.}}
\end{figure}

 The cone radius dependence of the inclusive jet cross section, $\sigr$, has
been studied. Measurements have been performed of the inclusive jet cross
section integrated above $\etjet>21$~GeV and $-0.5<\etajet<2$ for three
different cone radii ($\RRR$). These cross sections are given in the same $\q2$
and $W$ kinematic region as the measurements presented in Section~8.1.
As observed in the jet profiles (see Section~6), the uncertainties on the jet
cross sections due to a possible underlying event become reduced at
$\etjet>21$~GeV. The results for $\sigr$ are presented in
Figure~\ref{inclusive350} and Table~\ref{tabsec5}. The measured cross section is
consistent with a linear variation with $R$ in the range between 0.5 and 1.0.

 The results of LO and NLO QCD calculations of $\sigr$ \cite{kramer3}, which are
performed at the parton level, for different values of the renormalisation and
factorisation scales $\mu$ are shown in the inset of Figure~\ref{inclusive350}.
The LO and NLO GS96 (CTEQ4) sets of photon (proton) parton densities have been
used. The LO predictions do not depend on $R$ since there is only one parton
per jet and show a large variation with $\mu$. NLO calculations give the
lowest-non-trivial order $R$-dependent contributions to the jet cross section
and the $\mu$ dependence is largely reduced. However, at small (large) values
of $R$, the NLO predictions for $\sigr$ become a monotonically increasing
(decreasing) function of $\mu$. The calculations are most stable for
$R\approx 0.5-0.7$, consistent with the conclusions of \cite{sdellis}. The
uncertainty on the predicted cross section due to the choice of $\mu$,
estimated by changing $\mu$ from $\etjet/4$ to $\etjet$, is 5\% (20\%)
at $\r7$ ($\rr1$). 

 The slope of $\sigr$ depends on the choice of $\mu$, and is largest
(smallest) for small (large) values of $\mu$ (see inset of
Figure~\ref{inclusive350}). The slope of $\sigr$ in the NLO calculation with
$\mu=\etjet/4$ is closest to that of the measured cross section. In addition
to the uncertainty coming from the choice of $\mu$, the predictions are
affected by the value of $\rs$. QCD calculations with $\mu=\etjet/4$ and for
two values of $\rs$, $\rs=R$ and $2R$, are compared to the measurements in
Figure~\ref{inclusive350}. Since the LO predictions of the inclusive jet cross
section do not depend on $R$, the data show the need for QCD corrections. The
NLO calculations are consistent with the data within the theoretical and
experimental uncertainties, both of which are at the 20\% level.

\section{Summary and conclusions}

 Measurements of differential cross sections for inclusive jet photoproduction
in $e^+p$ collisions at HERA using the data collected by ZEUS have been 
presented. The cross sections refer to jets at the hadron level found with an 
iterative cone algorithm in the $\etaphi$ plane. Measurements of the jet 
cross sections with two different cone radii, $\R71$, have been performed. 
These cross sections are given in the kinematic region defined by 
$\q2 \leq 4$ \g2\ and \wr . 

 A comparison has been made of the transverse energy profiles around the jet
axis between data and the leading-logarithm parton-shower simulations of
PYTHIA. Requiring high $\etjet$ ($\etjet>21$~GeV) or using a cone radius 
of $\r7$ reduces the discrepancy between data and PYTHIA in the forward
region.

 NLO QCD calculations \cite{kramer3,harris1} using currently available
parametrisations of the photon parton distributions are compared to the
measured cross sections. The uncertainties on the calculations due to the 
choice of renormalisation and factorisation scales, and non-perturbative 
effects like the underlying event are smaller for jets with $\r7$ than in the
case of $\rr1$. The calculations describe the measured cross sections well for
jets defined with $\R71$ for $\etjet>$ 21 and 25~GeV. At lower values of
$\etjet$ differences between data and the calculations are seen in the
forward region for jets defined with $\rr1$. On the other hand, the
calculations describe well the measured differential cross sections in the
entire range of $\etajet$ for jets defined with $\r7$. These conclusions are
reinforced when the data are considered in different ranges of $W$. The 
uncertainties on the measurements with $\r7$ are comparable to the spread of 
the predictions using different parametrisations of the photon parton 
distributions. 

 The measured cross section for jets with $\etjet>21$~GeV and 
$-0.5<\etajet<2$ is consistent with a linear variation with the cone radius 
$R$ in the range between 0.5 and 1.0, and shows the need for QCD corrections.
The NLO calculations are consistent with the data within the 20\% theoretical
and experimental uncertainties.

\vspace{0.5cm}
\noindent {\Large\bf Acknowledgments}
\vspace{0.3cm}

 We thank the DESY Directorate for their strong support and encouragement. The
remarkable achievements of the HERA machine group were essential for the
successful completion of this work and are greatly appreciated. We would like
to thank B. Harris, M. Klasen, G. Kramer and J. Owens for providing us with 
their calculations.


\newpage
\clearpage

\begin{table}
\centering
\begin{tabular}{|c||c|c|}                \hline
$\eta^{jet}$ & $d\sigma/d\eta^{jet}\pm$ stat. $\pm$ syst. [pb] & syst.
$E^{jet}_T$-scale [pb]  \\ \hline\hline
 & $\etjet>14$ GeV &                          \\ \hline
{ -0.88} & $ 135\pm 15\pm  10$&$ (+50,-30)$   \\ \hline
{ -0.62} & $ 345\pm 25\pm  90$&$ (+80,-60)$   \\ \hline
{ -0.38} & $ 690\pm 35\pm  50$&$(+130,-110)$  \\ \hline
{ -0.12} & $1040\pm 40\pm 120$&$(+170,-120)$  \\ \hline
{  0.12} & $1330\pm 45\pm  90$&$(+190,-150)$  \\ \hline
{  0.38} & $1535\pm 45\pm 170$&$(+210,-160)$  \\ \hline
{  0.62} & $1790\pm 50\pm  60$&$(+220,-160)$  \\ \hline
{  0.88} & $1785\pm 50\pm  80$&$(+200,-160)$  \\ \hline
{  1.12} & $1715\pm 50\pm 110$&$(+170,-140)$  \\ \hline
{  1.38} & $1690\pm 50\pm  80$&$(+180,-170)$  \\ \hline
{  1.62} & $1655\pm 50\pm 110$&$(+230,-150)$  \\ \hline
{  1.88} & $1785\pm 50\pm 100$&$(+220,-210)$  \\ \hline\hline
 & $\etjet>17$ GeV &                         \\ \hline
{ -0.62} & $ 80\pm 10\pm 10$&$ (+30,-20)$  \\ \hline
{ -0.38} & $185\pm 15\pm 30$&$ (+50,-40)$  \\ \hline
{ -0.12} & $355\pm 25\pm 30$&$ (+70,-50)$  \\ \hline
{  0.12} & $500\pm 25\pm 30$&$ (+80,-70)$  \\ \hline
{  0.38} & $625\pm 30\pm 50$&$(+100,-90)$  \\ \hline
{  0.62} & $755\pm 35\pm 40$&$(+100,-70)$  \\ \hline
{  0.88} & $750\pm 35\pm 20$&$(+100,-80)$  \\ \hline
{  1.12} & $725\pm 35\pm 50$&$ (+80,-90)$  \\ \hline
{  1.38} & $690\pm 30\pm 40$&$ (+90,-70)$  \\ \hline
{  1.62} & $710\pm 35\pm 70$&$(+100,-90)$  \\ \hline
{  1.88} & $665\pm 30\pm 30$&$(+110,-70)$  \\ \hline\hline
 & $\etjet>21$ GeV &                          \\ \hline
{ -0.25} & $ 50\pm  5\pm 15$&$(+15,-10)$  \\ \hline
{  0.25} & $205\pm 10\pm 15$&$(+35,-35)$  \\ \hline
{  0.75} & $305\pm 15\pm  5$&$(+50,-40)$  \\ \hline
{  1.25} & $280\pm 15\pm 20$&$(+30,-35)$  \\ \hline
{  1.75} & $235\pm 15\pm 30$&$(+35,-35)$  \\ \hline\hline
 & $\etjet>25$ GeV &                          \\ \hline
{  0.25} & $ 70\pm  5\pm  5$&$(+15,-10)$  \\ \hline
{  0.75} & $125\pm 10\pm 15$&$(+20,-20)$  \\ \hline
{  1.25} & $125\pm 10\pm 15$&$(+20,-15)$  \\ \hline
{  1.75} & $120\pm 10\pm 10$&$(+25,-20)$  \\ \hline\hline
\end{tabular}
\caption{\label{tabsec1}{Differential $e^+p$ cross section $\seta$ for inclusive
jet production integrated above different $\etjet$ thresholds in the kinematic
region defined by $\q2\leq 4$~\g2\ and \wr\ for jets with a cone radius $\rr1$.
The statistical and systematic uncertainties $-$not associated with the
absolute energy scale of the jets$-$ are also indicated. The systematic
uncertainties associated to the absolute energy scale of the jets are quoted 
separately. The overall normalization uncertainty of 1.5\% is not included.}}
\end{table}

\newpage
\clearpage

\begin{table}
\centering
\begin{tabular}{|c||c|c|}                \hline
$\eta^{jet}$ & $d\sigma/d\eta^{jet}\pm$ stat. $\pm$ syst. [pb] & syst.
$E^{jet}_T$-scale [pb]  \\ \hline\hline
 & $\etjet>14$ GeV &                          \\ \hline
{ -0.88} & $  85\pm 15\pm 10$&$ (+30,-20)$  \\ \hline
{ -0.62} & $ 260\pm 20\pm 60$&$ (+70,-50)$  \\ \hline
{ -0.38} & $ 495\pm 30\pm 60$&$(+110,-80)$  \\ \hline
{ -0.12} & $ 735\pm 35\pm 80$&$(+120,-90)$  \\ \hline
{  0.12} & $ 935\pm 35\pm 50$&$(+120,-110)$  \\ \hline
{  0.38} & $1050\pm 40\pm 60$&$(+130,-120)$  \\ \hline
{  0.62} & $1225\pm 40\pm 60$&$(+140,-120)$  \\ \hline
{  0.88} & $1165\pm 40\pm 40$&$(+110,-110)$  \\ \hline
{  1.12} & $1065\pm 45\pm 70$&$(+120,-110)$  \\ \hline
{  1.38} & $1055\pm 40\pm 30$&$(+120,-100)$  \\ \hline
{  1.62} & $ 965\pm 35\pm 40$&$(+110,-100)$  \\ \hline
{  1.88} & $ 850\pm 35\pm 40$&$ (+90,-90)$  \\ \hline\hline
 & $\etjet>17$ GeV &                          \\ \hline
{ -0.62} & $ 60\pm 10\pm 10$&$ (+20,-20)$  \\ \hline
{ -0.38} & $150\pm 15\pm 20$&$ (+30,-30)$  \\ \hline
{ -0.12} & $240\pm 20\pm 30$&$ (+40,-40)$  \\ \hline
{  0.12} & $380\pm 25\pm 40$&$ (+60,-50)$  \\ \hline
{  0.38} & $470\pm 25\pm 10$&$ (+80,-70)$  \\ \hline
{  0.62} & $525\pm 25\pm 20$&$ (+70,-70)$  \\ \hline
{  0.88} & $560\pm 30\pm 40$&$ (+70,-70)$  \\ \hline
{  1.12} & $495\pm 30\pm 30$&$ (+50,-50)$  \\ \hline
{  1.38} & $465\pm 25\pm 30$&$ (+50,-50)$  \\ \hline
{  1.62} & $395\pm 25\pm 40$&$ (+50,-40)$  \\ \hline
{  1.88} & $385\pm 25\pm 30$&$ (+40,-50)$  \\ \hline\hline
 & $\etjet>21$ GeV &                          \\ \hline
{ -0.25} & $ 35\pm  5\pm 10$&$(+10,-10)$  \\ \hline
{  0.25} & $160\pm 10\pm  5$&$(+30,-25)$  \\ \hline
{  0.75} & $210\pm 10\pm  5$&$(+30,-25)$  \\ \hline
{  1.25} & $195\pm 10\pm 10$&$(+25,-25)$  \\ \hline
{  1.75} & $170\pm 10\pm 20$&$(+25,-25)$  \\ \hline\hline
 & $\etjet>25$ GeV &                          \\ \hline
{  0.25} & $64\pm 7\pm  6$&$(+10,-10)$  \\ \hline
{  0.75} & $93\pm 8\pm 11$&$(+15,-10)$  \\ \hline
{  1.25} & $85\pm 8\pm 15$&$(+15,-10)$  \\ \hline
{  1.75} & $85\pm 8\pm  8$&$(+15,-10)$  \\ \hline\hline
\end{tabular}
\caption{\label{tabsec2}{Differential $e^+p$ cross section $\seta$ for inclusive
jet production integrated above different $\etjet$ thresholds in the kinematic
region defined by $\q2\leq 4$~\g2\ and \wr\ for jets with a cone radius $\r7$.
Other details as in Table~\ref{tabsec1}.}}
\end{table}

\newpage
\clearpage

\begin{table}
\centering
\begin{tabular}{|c||c|c|}                \hline
$\eta^{jet}$ & $d\sigma/d\eta^{jet}\pm$ stat. $\pm$ syst. [pb] & syst.
$E^{jet}_T$-scale [pb]  \\ \hline\hline
 & \wrg\ &                          \\ \hline
{  0.25} & $340\pm 15\pm 70$&$ (+80,-50)$  \\ \hline
{  0.75} & $720\pm 25\pm 60$&$(+100,-80)$  \\ \hline
{  1.25} & $700\pm 25\pm 50$&$ (+80,-70)$  \\ \hline
{  1.75} & $725\pm 25\pm 50$&$ (+90,-80)$  \\ \hline\hline
 & \wrgg\ &                           \\ \hline
{ -0.25} & $305\pm 15\pm 60$&$ (+70,-50)$  \\ \hline
{  0.25} & $585\pm 20\pm 30$&$ (+70,-60)$  \\ \hline
{  0.75} & $555\pm 20\pm 50$&$ (+60,-40)$  \\ \hline
{  1.25} & $495\pm 20\pm 30$&$ (+50,-50)$  \\ \hline
{  1.75} & $500\pm 20\pm 40$&$ (+70,-50)$  \\ \hline\hline
 & \wrggg\ &                          \\ \hline
{ -0.75} & $220\pm 15\pm 30$&$ (+60,-40)$  \\ \hline
{ -0.25} & $535\pm 20\pm 30$&$ (+80,-60)$  \\ \hline
{  0.25} & $500\pm 20\pm 40$&$ (+50,-40)$  \\ \hline
{  0.75} & $495\pm 20\pm 30$&$ (+50,-40)$  \\ \hline
{  1.25} & $475\pm 20\pm 30$&$ (+50,-40)$  \\ \hline
{  1.75} & $475\pm 20\pm 20$&$ (+50,-50)$  \\ \hline\hline
\end{tabular}
\caption{\label{tabsec3}{Differential $e^+p$ cross section $\seta$ for inclusive
jet production integrated above $\etjet>14$~GeV in the kinematic region
defined by $\q2\leq 4$~\g2\ and in three regions of $W$ for jets with a cone
radius $\rr1$. Other details as in Table~\ref{tabsec1}.}}
\end{table}

\newpage
\clearpage

\begin{table}
\centering
\begin{tabular}{|c||c|c|}                \hline
$\eta^{jet}$ & $d\sigma/d\eta^{jet}\pm$ stat. $\pm$ syst. [pb] & syst.
$E^{jet}_T$-scale [pb]  \\ \hline\hline
 & \wrg\ &                        \\ \hline
{  0.25} & $230\pm 15\pm 30$&$ (+50,-40)$  \\ \hline
{  0.75} & $480\pm 20\pm 60$&$ (+60,-50)$  \\ \hline
{  1.25} & $450\pm 20\pm 30$&$ (+50,-50)$  \\ \hline
{  1.75} & $375\pm 15\pm 40$&$ (+50,-40)$  \\ \hline\hline
 & \wrgg\  &                       \\ \hline
{ -0.25} & $215\pm 15\pm 40$&$ (+60,-40)$  \\ \hline
{  0.25} & $400\pm 20\pm 10$&$ (+50,-40)$  \\ \hline
{  0.75} & $390\pm 15\pm 30$&$ (+40,-30)$  \\ \hline
{  1.25} & $320\pm 15\pm 30$&$ (+40,-30)$  \\ \hline
{  1.75} & $270\pm 15\pm 20$&$ (+30,-30)$  \\ \hline\hline
 & \wrggg\ &                           \\ \hline
{ -0.75} & $165\pm 15\pm 30$&$ (+50,-30)$  \\ \hline
{ -0.25} & $385\pm 15\pm 30$&$ (+50,-40)$  \\ \hline
{  0.25} & $365\pm 15\pm 10$&$ (+40,-40)$  \\ \hline
{  0.75} & $320\pm 15\pm 40$&$ (+30,-20)$  \\ \hline
{  1.25} & $285\pm 15\pm 10$&$ (+30,-30)$  \\ \hline
{  1.75} & $255\pm 15\pm 10$&$ (+20,-30)$  \\ \hline\hline
\end{tabular}
\caption{\label{tabsec4}{Differential $e^+p$ cross section $\seta$ for inclusive
jet production integrated above $\etjet>14$~GeV in the kinematic region
defined by $\q2\leq 4$~\g2\ and in three regions of $W$ for jets with a cone
radius $\r7$. Other details as in Table~\ref{tabsec1}.}}
\end{table}

\begin{table}
\centering
\begin{tabular}{|c||c|c|}                \hline
Cone radius & $\sigr \pm$ stat. $\pm$ syst. [pb] & syst. $E^{jet}_T$-scale
                                   [pb]  \\ \hline\hline
$R=0.5$ & $275\pm 10 \pm  30$&$ (+40,-30)$ \\ \hline
$R=0.7$ & $385\pm 10 \pm  20$&$ (+60,-50)$ \\ \hline
$R=1.0$ & $540\pm 15 \pm  40$&$ (+90,-80)$ \\  \hline
\end{tabular}
\caption{\label{tabsec5}{$e^+p$ cross section $\sigr$ for inclusive jet
production integrated above $\etjet>21$~GeV and $-0.5<\etajet <2$ in the
kinematic region defined by $\q2\leq 4$~\g2\ and \wr . Other details as in
Table~\ref{tabsec1}.}}
\end{table}

\newpage
\clearpage
\parskip 0mm

\newpage
\clearpage
\parskip 0mm

\newpage
\clearpage
\parskip 0mm

\newpage
\clearpage
\parskip 0mm

\newpage
\clearpage
\parskip 0mm

\newpage
\clearpage
\parskip 0mm

\newpage
\clearpage
\parskip 0mm

\newpage
\clearpage
\parskip 0mm

\end{document}